\documentclass[iop]{emulateapj}
\usepackage{epsfig}
\usepackage{amsfonts,amssymb,amsmath}
  
\begin{document}

\title{The accretion of dark matter subhaloes within the cosmic web: primordial anisotropic distribution and its universality}
\author{Xi Kang$^{1}$ \& Peng Wang$^{1,2}$}
\affil{$^1$Purple Mountain Observatory, the Partner Group of MPI f\"ur Astronomie, 2 West Beijing Road, Nanjing 210008, China\\
        $^2$Graduate School, University of the Chinese Academy of Science, 19A, Yuquan Road, Beijing 100049, China}

\email{kangxi@pmo.ac.cn}

\begin{abstract}

The distribution  of galaxies displays anisotropy  on different scales
and  it is  often referred  as  galaxy alignment.   To understand  the
origin of galaxy alignments on  small scales, one must investigate how
galaxies  were  accreted in  the  early  universe and  quantify  their
primordial anisotropic at the time of  accretion. In this paper we use
N-body simulations to investigate the accretion of subhaloes, focusing
on  their  alignment with  halo  shape  and  the orientation  of  mass
distribution on large  scale, defined using the hessian  matrix of the
density  field.  The  large/small (e1/e3)  eigenvalues of  the hessian
matrix  define the  fast/slow collapse  direction of  matter on  large
scale.  We find that: 1) the halo  major axis is well aligned with the
e3 (slow collapse)  direction, and it is stronger  for massive haloes;
2) subhaloes  are predominately accreted  along the major axis  of the
host halo, and the alignment increases  with the host halo mass.  Most
importantly, this  alignment is  universal; 3) accretion  of subhaloes
with respect to the e3  direction is not universal. In  massive haloes, subhaloes
are accreted along the e3 (even stronger than the alignment with the halo major
axis), but in low-mass haloes  subhaloes are accreted perpendicular to
e3. The transit  mass is lower at high redshift.  The last result well
explains  the puzzled  correlation  (both in  recent observations  and
simulations   )  that   massive   galaxies/haloes   have  their   spin
perpendicular   to   the   filament,   and  the   spin   of   low-mass
galaxies/haloes  is  slightly aligned  with  the  filament, under  the
assumption that the orbital angular momentum of subhaloes is converted
to halo spin.

\end{abstract}



\section{Introduction}
\label{sec:intro}

In the  cold dark  matter universe (CDM),  structures emerge  from the
initial seed of perturbations  via gravitational instability. On large
scales,  the   mass  distribution  is  characterized   as  cosmic  web
(filaments, sheets, clusters, voids), which can be fairly described by
a linear  theory and  the Zel'dovich approximation  (Zel'dovich 1970).
On small scales, dark matter haloes  form by collapse of matter at the
intersection  of filaments,  and they  gradually merge  and grow  in a
hierarchical and non-linear  way.  The detail of the  dark matter halo
properties and  mass distribution  on large  scales can  be accurately
studied by  N-body simulations (e.g.   Springel et al.   2006).  Under
such  a  formation  scenario,  it   is  naturally  expected  that  the
properties  of haloes  are correlated  with the  large-scale structure
(LSS).  For  example, the halo shape  is well aligned with  the cosmic
filament  (e.g.,  van  Haarlen  \&  van  de  Weygaert  1993;  Hahn  et
al.  2007a; Aragon-Calvo  et  al. 2007a;  Zhang  et
al. 2009;  for a summary,  see Forero-Romero et  al. 2014 ),  the halo
spin  is  originated   from  the  tidal  torque  force   by  the  mass
distribution on large  scales (White 1984), and its  direction is well
correlated  with  the  cosmic  web  (e.g.,  van  de  Weygaert  \&
  Bertschinger 1996; Bond  et al. 1996; Lee 2004;  Bailin \& Steinmetz
  2005;  Hahn et  al.  2007a;  Zhang et  al.   2009;  Libeskind et  al.
  2012). Most importantly, the alignment between halo spin and the LSS
  has a dependence on halo mass  (e.g., Aragon-Calvo et al. 2007a; Hahn
  et al.  2007b; Codis et  al. 2012; Trowland  et al. 2013;  Dubois et
  al. 2014; Zhang et al. 2015).

These theoretically predicted correlations of the halo properties with
LSS are manifested by the  distribution of galaxies inside dark matter
haloes, as galaxies are 'luminous'  tracers of merged haloes formed at
early times. For  example, satellite galaxies are found  to be aligned
with the  major axis  of central galaxies,  with strong  dependence on
galaxy colors and  halo mass (Brainerd 2005; Yang et  al. 2006). Local
dwarf  satellites   in  the   Milky  Way   and  M31   are  distributed
anisotropically and most are in a large thin disk (Kroupa et al. 2005;
Metz et  al.  2007;  Ibata et  al. 2013). The  shapes of  luminous red
galaxies are correlated with each other up to very large scales (e.g.,
Faltenbachler et  al. 2007;  Okumura et  al. 2009;  Li et  al.  2013).
Galaxy  spin  is  also  found  to  correlate  with  LSS  although  the
measurements  of  the  galaxy  spin are  not  straightforward  (e.g.,
  Trujillo et al.  2006; Jones et al. 2010).  More interestingly,
    it has been recently found that the correlation of the galaxy spin
    with the LSS  depends on galaxy  type/mass, such that the  spin of
    early type galaxies  is perpendicular to the  filament around, but
    late-type  galaxies  have  their  spin slightly  parallel  to  the
    filament (Tempel \& Libeskind 2013; Zhang et al. 2015), in amazing
    good   agreement   with   the  theoretical   expectations   (e.g.,
    Aragon-Calvo et al. 2007a).

The observed anisotropic distribution  of galaxies on different scales
raises  the question  of its  origin.  Some studies  (e.g., Benson  et
al. 2005;  Wang et al.  2005; Libeskind et  al. 2005) have  shown that
satellite galaxies  are accreted anisotropically during  the formation
of dark matter haloes, and this primordial anisotropy is the origin of
the current anisotropic distribution  of satellite galaxies.  However,
the  observed  distribution  of  galaxies within  DM  haloes  involves
complicated baryonic  processes (e.g., Kang et  al. 2005; Vogelsberger
et al.  2014) and galaxies  have experienced the  non-linear dynamical
evolution inside  dark matter  haloes.  Along with  the fact  that the
signal of  the galaxy anisotropic  distribution is much weaker,  it is
therefore  difficult  to infer  how  much  of the  currently  observed
anisotropic distribution of galaxies is from the primordial anisotropy
set on large scales before galaxies  were accreted, and how much of it
is from  the non-linear  evolution inside the  dark matter  halo.  For
example, the anisotropic distribution of satellites respect to central
galaxies can  be purely  ascribed to the  non-linear evolution  or the
non-spherical nature  of dark matter  haloes (e.g.,Jing \&  Suto 2002;
Kang et al. 2007;  Wang et al., 2014; Dong et  al. 2014; Debattista et
al. 2015). For  a recent review on galaxy alignments  and its relation
with the halo shape  or the LSS, see Joachimi et  al. (2015). The
  flip  of  the  halo   spin-LSS  correlation  (e.g.  Aragon-Calvo  et
  al.   2007a)   also   attracts  great   attention   recently.   Using
  hydrodynamical simulations, it was found that the flip is related to
  the halo merger history and  cold gas accretion during the formation
  of elliptical and  spiral galaxies (e.g., Codis et  al. 2012; Welker
  et al. 2014; Codis et al. 2015).

To understand the observed correlation of the galaxy distribution with
respect to  the halo shape  or the LSS in  detail, one needs  to trace
galaxies back to  the time of their accretion, to  quantify the degree
of the primordial correlation with DM  haloes or the LSS at that time.
This is  often done  by tracing  subhaloes back  to early  times using
N-body simulations.  In this paper  we study the spatial  alignment of
subhaloes with respect  to the host halo  shape and the LSS,  and in a
future paper we  will focus on the velocity/spin  correlation with LSS
at the  time of accretion.  There  were only a few  studies using this
kind  of approach.  Benson  et  al. (2005)  found  that subhaloes  are
preferentially accreted in the planes  defined by the major and middle
axes of the host haloes. Wang  et al. (2005) also found that subhaloes
are  accreted along  halo  major  axis and  massive  subhaloes show  a
stronger trend.  However, they both  ignored the correlation  with the
LSS. Recently,  Libeskind et al.  (2014)  and Shi et al.   (2015) find
that  subhaloes are  predominantly  accreted along  the filament,  but
their classification of cosmic web is based on velocity shear or tidal
field. In our paper we classify the cosmic web by means of the density
field in order to classify the  LSS environment, and our analysis goes
to very low-mass haloes. In particular,  we will show that our results
are helpful to explain the puzzled non-universality in the correlation
between halo/galaxy spin with the LSS.
 
 The  paper is  organized as  follows. In  Section.2 we  introduce the
 simulations  and the  methods  to  quantify the  halo  shape and  the
 LSS. In section.3  we show how the halo shape  is correlated with the
 LSS  and  the anisotropic  distribution  of  accreted subhaloes  with
 respect  to the  halo  shape  and the  LSS.  Then,  we present  their
 dependence on halo  mass and redshift. In section.4  we summarize our
 results  and briefly  discuss how  our  results help  to explain  the
 correlation between the halo spin and the LSS.

\section{N-body simulation and LSS classification}
\label{sec:methods}

In this work  we use two N-body simulations with  different box sizes.
Both simulations are run with  the Gadget-2 code (Springel 2005) using
the same number  of particles of $1024^{3}$ and  the same cosmological
parameters  from  the  WMAP7  data   (Komatsu  et  al.  2011,  namely:
$\Omega_{\Lambda}=0.73$,      $\Omega_{m}=0.27$,      $h=0.7$      and
$\sigma_{8}=0.81$.  The  box sizes  are $200Mpc/h$ and  $65Mpc/h$, and
the particle  masses in  these two  simulations are  $5.5\times 10^{8}
M_{\odot}/h$, and  $1.8\times 10^{7}M_{\odot}/h$,  respectively.  With
the larger  simulation box we  get more well resolved  massive haloes,
while the  smaller box  one allows  us to resolve  haloes down  to low
masses,  about  $5\times  10^{9}M_{\odot}/h$.   We note  that  in  the
following   analysis  the   results   for  haloes   with  mass   below
$10^{12}M_{\odot}$  are  from  the  simulation  with  small  box  size
($65Mpc/h$),  and for  massive haloes  we  use the  results from  the
  $200Mpc/h$ simulation.

From  the simulation  output we  firstly identify  dark matter  haloes
using the standard friend-of-friend (FOF) algorithm, and each FOF halo
should contain  at least  20 particles.  Then,  subhaloes in  each FOF
halo are found  using the SUBFIND algorithm (Springel et al. 2001)  and their merger
trees are also constructed. For details  the readers can refer to Kang
et al. (2005). We trace each subhalo back to the time when it was last
identified as  an independent  FOF halo  (labeled as  $h$). In  a very
short time  period (the next snapshot)  the small halo $h$  will merge
with a larger halo (host halo,  labeled as $H$). By means of positions
and velocities of the haloes $h$ and  $H$, we can obtain the time and 
position when the halo $h$ crosses the virial radius of halo $H$ for 
the first  time. Note that  in the following,  when we refer  to the 
subhalo  distribution at  accretion,  we are  actually referring  to 
either the  distribution of the  halo $h$  with respect to  the host 
halo $H$, or  with the large scale  around the halo $H$  at the time 
when halo $h$ crosses the virial radius of $H$ for the first time.

For each  FOF halo  its virial  mass is the  enclosed mass  within the
virial radius inside  which the mean density  is $\Delta_{c}(z)$ times
the  critical density  of  the  Universe (Bryan  \&  Norman 1998).  To
measure the  shape of each halo,  we follow the traditional  method to
calculate  the  normalized  inertia   tensor  defined  as  (Bailin  \&
Steinmetz 2005),
\begin{equation}
I_{ij} = \Sigma_{n} \frac {x_{i,n}x_{j,n}}{R^{2}_{n}},
\end{equation}
where $x_{i,n}$ is  the distance component of the particle  $n$ to the
halo center, and the summation is over all particles inside the virial
radius. The three  eigenvectors of $I_{ij}$ define  the orientation of
the three axes of a halo,  and the direction of the largest eigenvalue
defines the major  axis. Note that the minimum number  of particles in
each  host halo  $H$  is taken  as  500,  so the  halo  shape is  more
accurately determined  (Jing \&  Suto 2002). We also  tested the halo 
shapes (the three eigenvectoes)  using all particles of the FOF halo, and  
found that they  agree quite  good with those  from those using the particles 
inside the virial radius.
 
\begin{figure*}
\centerline{\psfig{figure=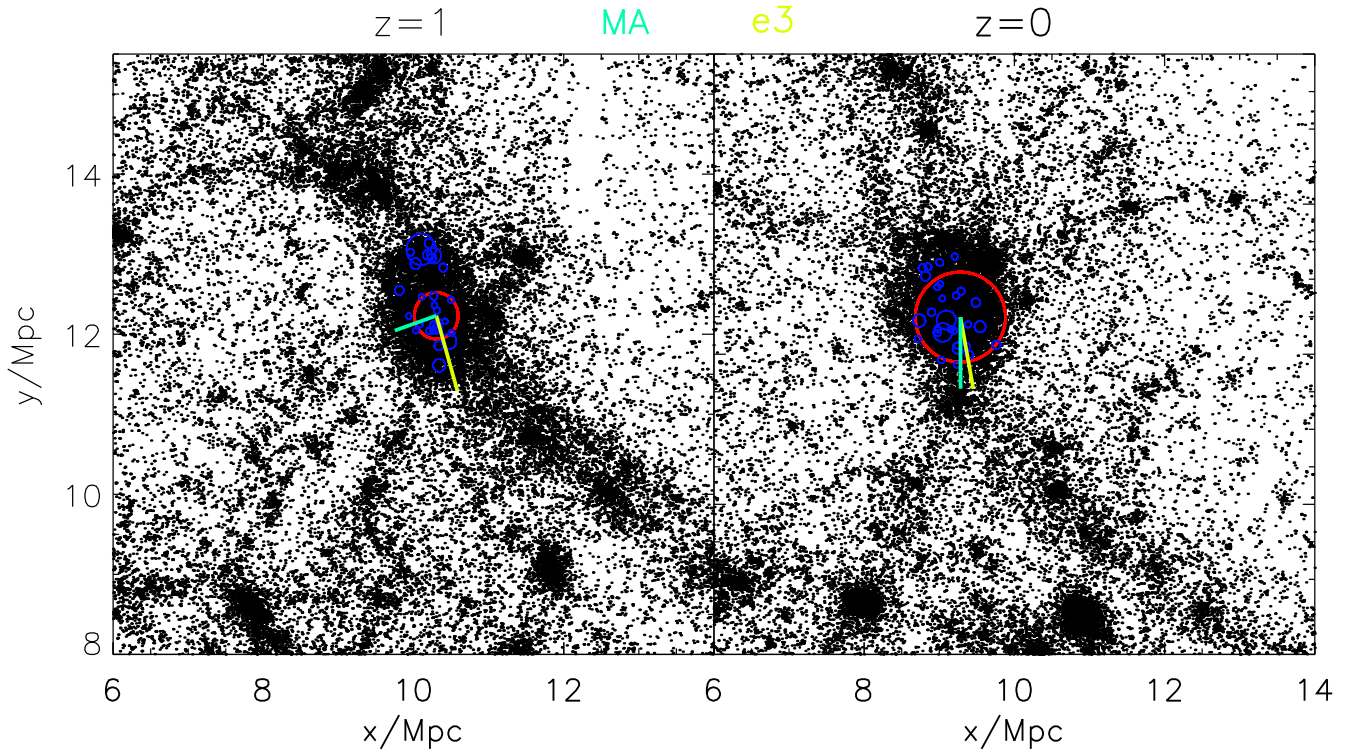,width=0.98\textwidth}}
\caption{Slices  from  the  simulation. Left:  the 2D mass  distribution
  around a halo at $z=1$, the red circle denotes its virial radius, and
  blue circles denote those subhaloes to  be merged with the host halo
  in a short time (some subhaloes are located inside the red circle due to projection effect). 
  Yellow and cyan  lines show the e3 direction of the
  density  field  on   large  scales,  and  the  major   axis  of  the
  halo. Right: the slice at $z=0$,  the halo is the descendent of that
  shown  in  the   left  panel,  and  blue  circles   show  the  $z=0$
  distributions  of  those accreted  subhaloes  labelled  in the  left
  panel.}
\label{fig:slice}
\end{figure*}

\begin{figure*}
\centerline{\psfig{figure=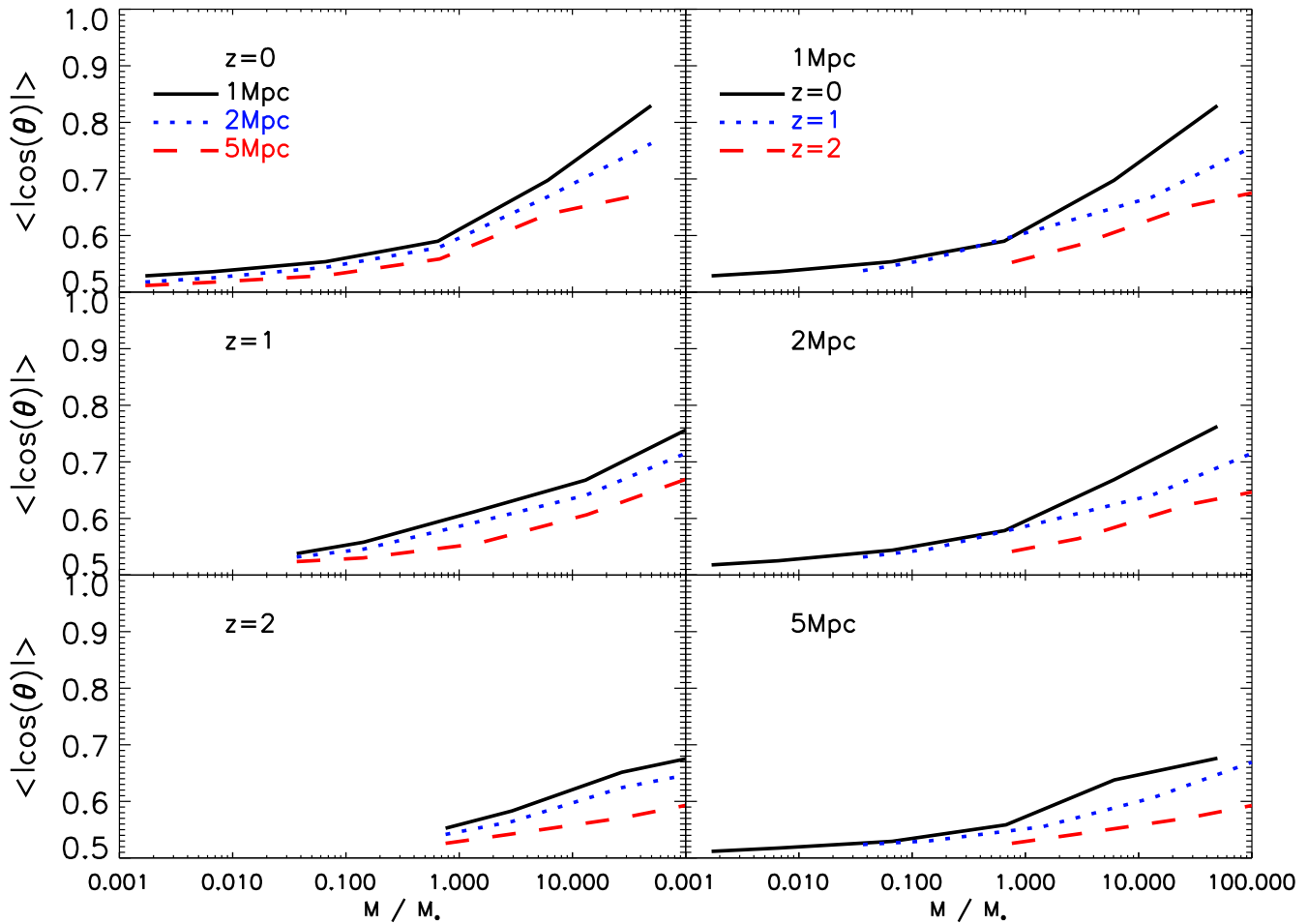,width=0.98\textwidth}}
\caption{The mean  angle between the  halo major axis and  the slowest
  collapse  direction on  large scales,  $e_{3}$, calculated  from the
  Hessian matrix using different smooth  length.  The left panels show
  the  dependence of  the  alignment  on the  smooth  length at  given
  redshift, and  the right panels  show the redshift evolution  of the
  alignment for a given smooth length.}
\label{fig:haloMA-e3}
\end{figure*}

To define the LSS environment  around each halo, we follow
the  most used  method (e.g.,  Hahn et al. 2007a; Arag'on-Calvo  et al.  2007b; Zhang  et
al. 2009,) by calculating the Hessian  matrix at  the position  of each
halo. The Hessian matrix of the smoothed mass density field is defined
as,
\begin{equation}
H_{ij}=\frac{\partial^{2}\rho_{s}(x)} {\partial x_{i}\partial x_{j}}
\end{equation}
where  $\rho_{s}(x)$  is  the  smoothed  density  with  smooth  length
$r_{s}$. There is lack of systematic census on which smooth length is
the best  to characterize the  mass distribution on large  scales (for
partial  discussions,   see  Hahn  et  al.   2007b;  Forero-Romero  et
al. 2014).  In most  studies, a  constant smooth  length ($0.5  \sim 2
Mpc/h$) is used at $z=0$ (e.g., Hahn  et al. 2007a; Zhang et al. 2009;
Codis et al. 2012; Trowland et al. 2013). However, it is found that an
evolving smooth length  is more physical reasonable and  can better 
describe  the  evolution  of  the   LSS  environment  (e.g.,  Hahn  et
al. 2007b; Libeskind  et al. 2014). Following Hahn et  al. (2007b), we
adopt  a smooth  length roughly  scaled as  the virial  radius of  the
typical mass for halo collapse at different redshfit (often written as
$M_{*}$). To be able to compare  with other studies at $z=0$, we adopt
$r_{s} \sim 2/(1+z)  Mpc/h$. Actually, we also found  that our results
are not significantly affected by using  other constant smooth length (1, 2, 5
Mpc/h)   at  all   redshifts.  We   will  show   some  comparison   in
Fig.~\ref{fig:haloMA-e3}.

The eigenvalues  ($\lambda_{1}>\lambda_{2}>\lambda_{3}$) of the Hessian matrix define 
the LSS environment around a halo, and it is
classified as $cluster,  filament, sheet$ and $void$  depending on the
number  of  positive eigenvalues.  For  example,  there is  no  positive
eigenvalue for  a $cluster$, one  positive eigenvalue for  a $filament$, and
two positives for  a $sheet$. This  classification  of  halo  environment   well  mimics  the
description  of  Zel'dovich  theory  (for  a  review,  see  Cautun  et
al.  2014). Basically,  the eigenvector  $e_{1}$ (corresponding  to the
largest eigenvalue $\lambda_{1}$) defines the fast collapse direction,
and  $e_{3}$  corresponds  to  the  slowest  collapse  direction.  For
example,  for a  halo in  a  filament environment,  $e_{3}$ gives  the
direction  where the  matter has  not  collapsed on  large scales,  and
future mass accretion will mostly happen along $e_{3}$.  As the matter
distribution along  $e_{3}$  has not  strongly collapsed, the  halo is
less compressed along this direction, and  this is why  the halo major axis 
will then have  a tendency to align with $e_{3}$,  which has been shown in
many studies (e.g., van Harrlem \& van de Weygaert 1993).

  Fig.~\ref{fig:slice}   illustrates  the   space   configuration
  described above. The left panel shows the mass distribution around a
  selected host halo at $z=1$, with  virial radius labelled as the red
  circle. The blue circles denote  the subhaloes which will merge with
  the host  halo in  a short  time scales. The  yellow line  shows the
  $e_{3}$ direction,  which is seen to  be well aligned with  the mass
  distribution on large  scales. Cyan line is the  direction along the
  major axis of  the host halo. The position angles  of subhaloes with
  the   cyan/yellow  line   are  $\theta_{MA}$   and  $\theta_{e_{3}}$
  respectively.  The  right  panel  shows the  distribution  of  those
  accreted  subhalo  at  $z=0$, their  anisotropic  distribution  with
  respect to the host major axis can also be seen in the plot.

 With  the above  descriptions,  we  are able  to  derive the  spatial
 distribution of subhaloes  with respect to the host halo  and its LSS
 environment around.  The angular position of the halo $h$ at the time
 of  accretion with  respect to  the  major axis  of the  halo $H$  is
 labeled  as  $\theta_{MA}$, and  the  angle  respect to  the  slowest
 collapse  direction   ($e_{3}$)  centered   on  $H$  is   labeled  as
 $\theta_{e_{3}}$. For  a random  distribution, the expected  value of
 $<|cos(\theta)|>$ is  0.5, and if  $<|cos (\theta)|>$ is  larger than
 0.5, we refer to  it as an alignment with the halo  major axis or the
 LSS.

\section{Results}

\subsection{Accretion along halo major axis and the LSS}

Observations of  galaxy  distributions have found that  satellite galaxies 
are  aligned with  the major  axes of  their host  galaxies, and  this
alignment  is  dependent   on  the mass/color  of  the  hosts   (e.g.,  Yang  et
al. 2006).  Due to the fact that the observed signal is  weak compared to  that from
N-body simulations  (e.g., Kang et  al. 2007), satellite galaxies  alignments 
can be  explained if the central  galaxy roughly follows the  shape of
the dark matter halo and the tracing becomes stronger with halo mass
(e.g.,  Dong  et al.  2014).  Such  an  explanation implies  that  the
observed alignment can be purely  ascribed to the non-linear evolution
inside DM  haloes.

 In the first  part of this section we study  if such an alignment
  between subhaloes positions and halo major  axis (or $e_{3}$ of the LSS) 
can be  seen at the  time of  accretion,  and  investigate which  kind  of  alignment is  stronger
  ($\theta_{MA}$ or $\theta_{e_{3}}$).  Before discussing our results,
  we first  check how the  shape of  dark matter haloes  is correlated
  with the  tidal field  around them.  In  Fig.~\ref{fig:haloMA-e3} we
  plot  the average  angle  between the  major axis  of  the halo  and
  $e_{3}$, the slowest  collapse direction of the mass on  LSS. In the
  left panel, we show the dependence of the alignment on smooth length
  ($1,  2, 5  Mpc/h$), and  in the  right panel  we show  the redshift
  evolution for a fixed smooth length. Note that here the halo mass is
  scaled by  the characteristic mass  ($M_{*}$) of halo  collapsing at
  given  different. By  scaling  the  halo mass  with  $M_{*}$ we  are
  actually studying the peak of  the density field (with given height)
  with its  LSS around,  which has  a more  physical meaning  (Hahn et
  al. 2007b; Trowland et al. 2013;)

\begin{figure*}
\centerline{\psfig{figure=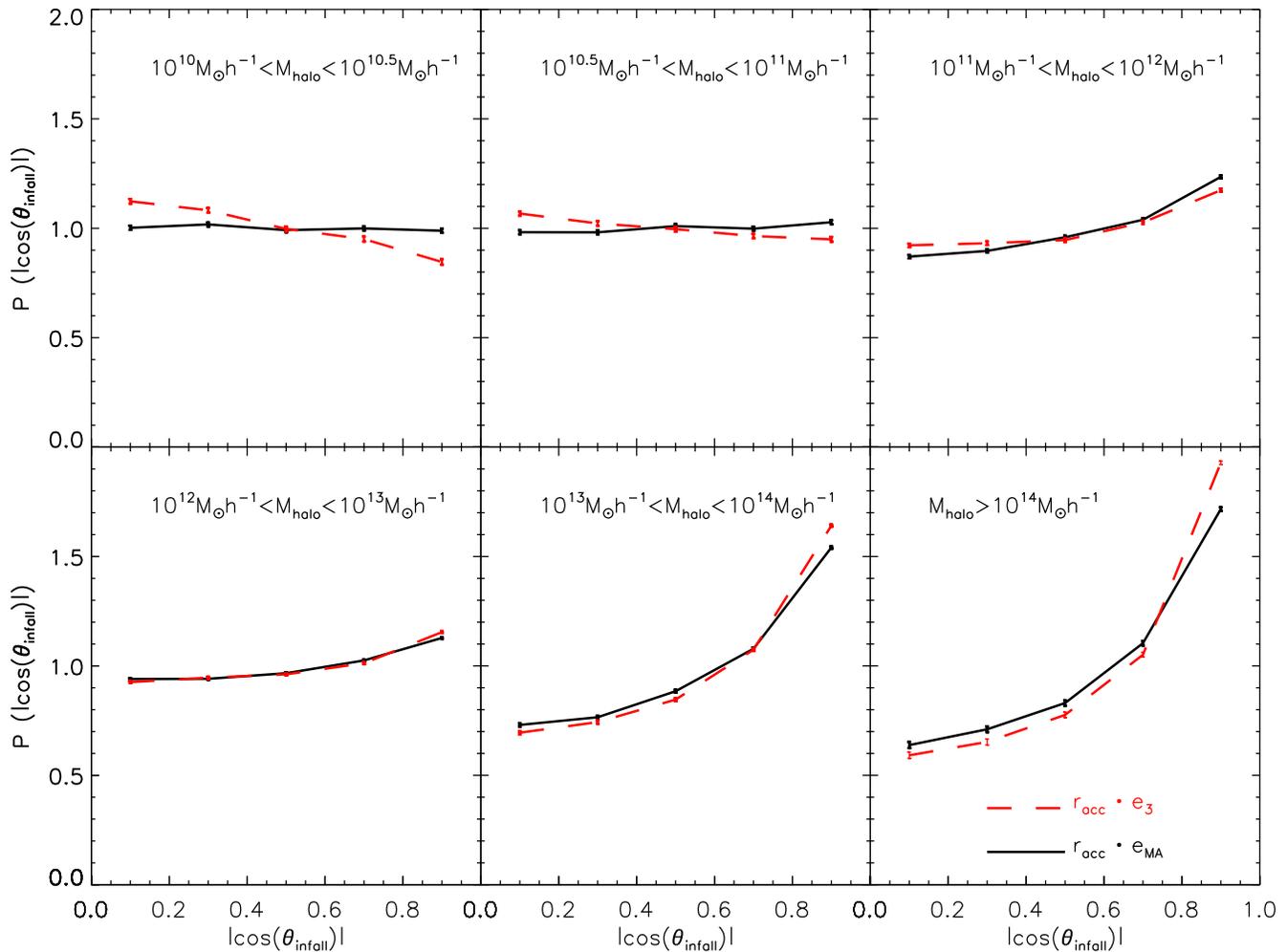,width=0.98\textwidth}}
\caption{The alignment of  subhaloes with the halo major  axis and the
  $e_{3}$ direction of  the LSS at their time of  infall (see text for
  detail). Different  panels are  for the final  host haloes  at $z=0$
  with different mass.  The alignment with the halo shape (black solid
  lines) is universal and increases with halo mass. The alignment with
  $e_{3}$ (red dashed lines) is stronger  in massive haloes, but it is
  reversed for low-mass haloes where it is perpendicular to $e_{3}$.}
\label{fig:infall1}
\end{figure*}

 Fig.~\ref{fig:haloMA-e3} shows that the major axis of the DM halo
  is well aligned with its  LSS ($e_{3}$), and the alignment increases
  with halo  mass. The left  panel shows  that at given  redshift, the
  alignment  is  stronger  for   a  smaller  smooth  length  (1Mpc/h),
  indicating  that the  mass  distribution at  large  scale is  better
  described by  this smooth length.  The right  panel shows that  at a
  given smooth  length, the alignment  is evolving with  redshift, and
  being stronger at low redshift  due to the non-linear evolution. The
  universal alignment between halo major  axis with the orientation of
  the density field  ($e_{3}$) agrees well with  the published results
  so   far  (e.g.,   Forero-Romero   et  al.   2014,  and   references
  therein). The results clearly show  that when a halo collapses
  its  internal  mass  will  be strongly  compressed  along  the  fast
  collapse direction  where the mass  over-density is highest,  so the
  particles will more likely move along the least compressed direction
  $e_{3}$.  Such a  scenario well  captures the  spirit of  Zel'dovich
  theory and is a manifestation  of the correlation between halo shape
  (density peak)  and the  orientation of the  density field  on large
  scale (van de  Weygaert \& Bertschinger 1996; Bond et  al. 1996; Lee
  \&  Pen  2000;  Porciani  et  al. 2002;  Rossi  2013;  Libeskind  et
  al. 2014).

 Now we investigate the primordial anisotropy of subhaloes at the time
 of accretion. As  found in observations, the  alignment of satellites
 is stronger in massive host  haloes.  For this reason, we investigate
 if the  primordial anisotropy is also  dependent on the mass  of host
 haloes. We select host haloes at $z=0$ with a wide range of mass from
 $10^{10}M_{\odot}$ to  $10^{14}M_{\odot}$, and trace  their subhaloes
 back to  the accretion times.  For  each accretion event, we  can get
 the angle between the infalling position  of a subhalo (halo $h$) and
 the major  axis of the  host halo, $cos(\theta_{MA})$, and  the angle
 with the $e_{3}$ direction, $cos(\theta_{e_{3}})$, respectively.  The
 distributions  of  them  are  shown  in  Fig.~\ref{fig:infall1},  and
 different panels  are for host  haloes with different virial  mass at
 $z=0$.    The    black    solid   lines    are    distributions    of
 $|cos(\theta_{MA})|$    and     red    dashed    lines     are    for
 $|cos(\theta_{e_{3}})|$.

 Some  interesting results  can be  seen from  Fig.~\ref{fig:infall1}.
 Firstly, the infalling  subhaloes have a tendency to  align with halo
 major axes, and the alignment increases with host halo mass. However,
 the alignment with $e_{3}$ is  different and, more importantly, it is
 not universal.  In massive haloes  (lower right panel), the alignment
 is even stronger than the halo  major axis, but it is anti-aligned in
 low-mass host haloes (upper left  panels), where subhaloes are accreted
 perpendicularly to $e_{3}$.  So, the  accretion in low-mass haloes is
 preferentially along the fast  collapse direction. This has important
 implication in explaining why the  spin of low-mass haloes is aligned
 with  the   filament.  We  will   show  the  results  in   detail  in
 Section.~\ref{sec:dependence}.

\begin{figure*}
\centerline{\psfig{figure=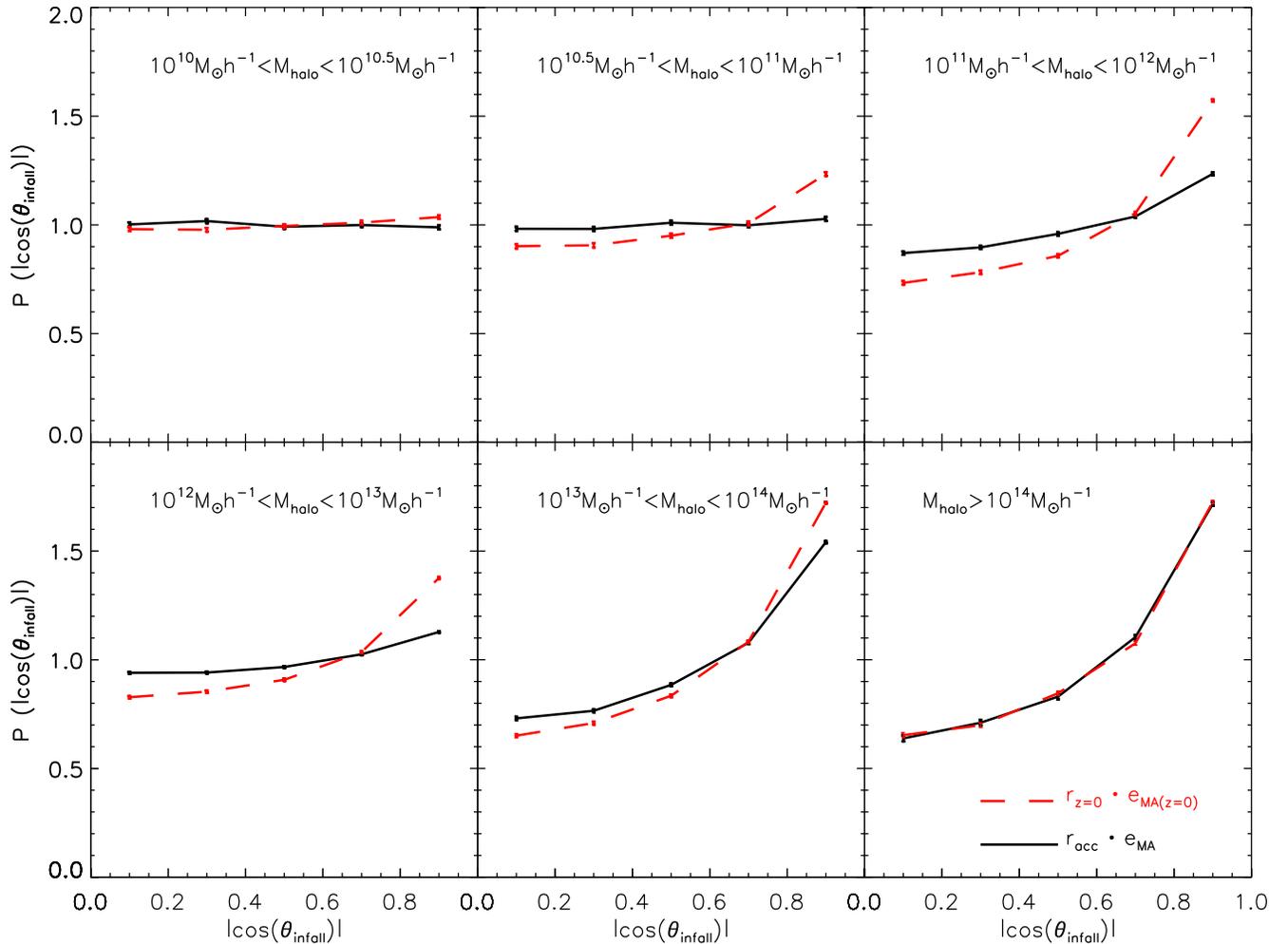,width=0.98\textwidth}}
\caption{The current alignment of subhaloes with the halo major axis at z=0 (red dashed lines), and the alignment at accretion (black solid lines). It is found that the current alignment with the halo major axis is higher than that in the past, indicating a non-linear evolution effect which is more obvious in intermediate halo masses.}
\label{fig:infall2}
\end{figure*} 

 The  Zel'dovich theory  predicts  that  structure formation  proceeds
 subsequently along  the three  eigenvectors of  LSS, i.e,  it firstly
 collapse along $e_{1}$ to form a  sheet, and then the second collapse
 happens along  $e_{2}$ to form  a filament. A  halo will form  if the
 matter along  the filament ($e_{3}$)  continues to collapse.  In
   this secnario, the accretion of a growing halo is expected to along
   the recent  collapsed direction ($e_{3}$). Although  the Zel'dovich
   approximation is only valid on linear scale in Lagrangian space, it
   remains valid  to describe the  feature in the  Eulerian non-linear
   regime, as  we see in the  lower panels. Only on  very small scales
   the accretion departs from the predictions of the Zel'dovich theory
   (upper left panels).  On small scales, the mass  accretion is more
   randomly,  and being  slightly aligned  with the  halo major  axis,
   implying that mass accretion is more likely determined by the local halo 
   potential.  Departure  of anisotropic  accretion  from  the
   Zel'dovich prediction has been realized for a long time (e.g., Icke
   1973; Silk \&  White 1978; Eisenstein \& Loeb 1996).  Bond \& Myers
   (1996)  provided a  comprehensive description  about the  nonlinear
   anisotropic  collapse  of  ellipsoids,  which  matches  better  the
   simulation results than the Zel'dovich theory.

 Our  results about  subhalo accretion  with $e_{3}$  marginally agree
 with  the recent  results  of  Libeskind et  al.  (2014)  and Shi  et
 al. (2015),  who have  also studied the  accretion of  subhaloes with
 respect to  the LSS.  Libeskind  et al.  (2014) found  that subhaloes
 are  predominantly accreted  along  the $e_{3}$  direction, and  this
 effect increases with  both the subhalo and host halo  masses.  As we
 will  see in  the next  section, they  obtained a  positive alignment
 because their bins of host halo mass  is wider, thus the anti-alignment
 from  low-mass  haloes  is  immersed in  the positive alignment  signal  from
 high-mass haloes.  Shi et al. (2015)  use a method similar to ours to
 define the large scale structure,  and they also found that subhaloes
 are accreted along the $e_{3}$ direction.  However, their analysis is
 only for  massive haloes  ($>10^{12}M_{\odot}$) where their results are consistent with ours. 
It is not clear what are the results for
 subhalo accretion in the low-mass haloes in Shi et al. (2015).

In  Fig.~\ref{fig:infall2}   we  compare  the  current   alignment  of
subhaloes (using subhaloes' positions and the host halo major axes at  $z=0$) 
with that  at infall. The black solid lines  are the same as in  Fig.~\ref{fig:infall1}, and the
red dashed ones show the current alignment  in different host haloes. It is seen that the $z=0$  alignment  is  stronger in  massive  haloes, which  is
consistent  with what the data suggests.   In general, the  current
alignment of subhaloes with their  hosts is larger than the alignment
at  the time of  accretion.  A stronger evolution is more obvious in  haloes with mass around $10^{12}M_{\odot}$.  However,  observed galaxies often
live  in  haloes  with  mass larger  than  $10^{12}M_{\odot}$ (Yang  et
al.  2006), and  the observed  alignment of  galaxies is much weaker, even compared with  the primordial alignment (black  solid  lines). So, it
is still difficult to  quantify how much of the observed signal is from the
primordial alignment  and from the evolution.

\begin{figure*}
\centerline{\psfig{figure=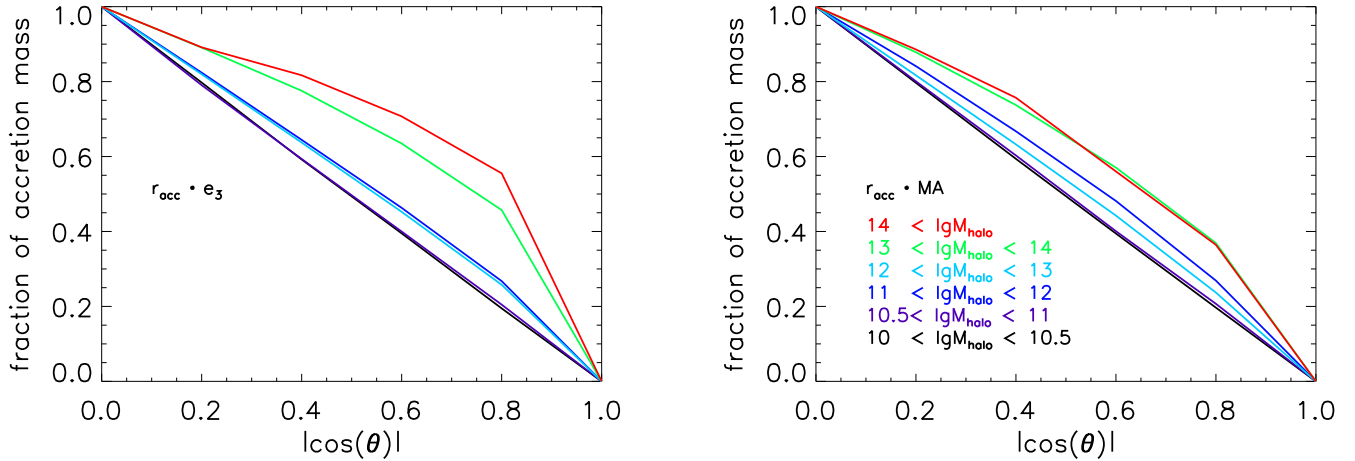,width=0.98\textwidth}}
\caption{The cumulative fraction of accreted mass in subhaloes with  a given alignment angle. Left panel is for alignment with $e_{3}$, and right for alignment with halo major axis.  A higher fraction of  mass is accreted along $e_{3}$ in massive haloes, and it becomes almost isotropic in low-mass haloes.}
\label{fig:infall3}
\end{figure*}

\begin{figure*}
\centerline{\psfig{figure=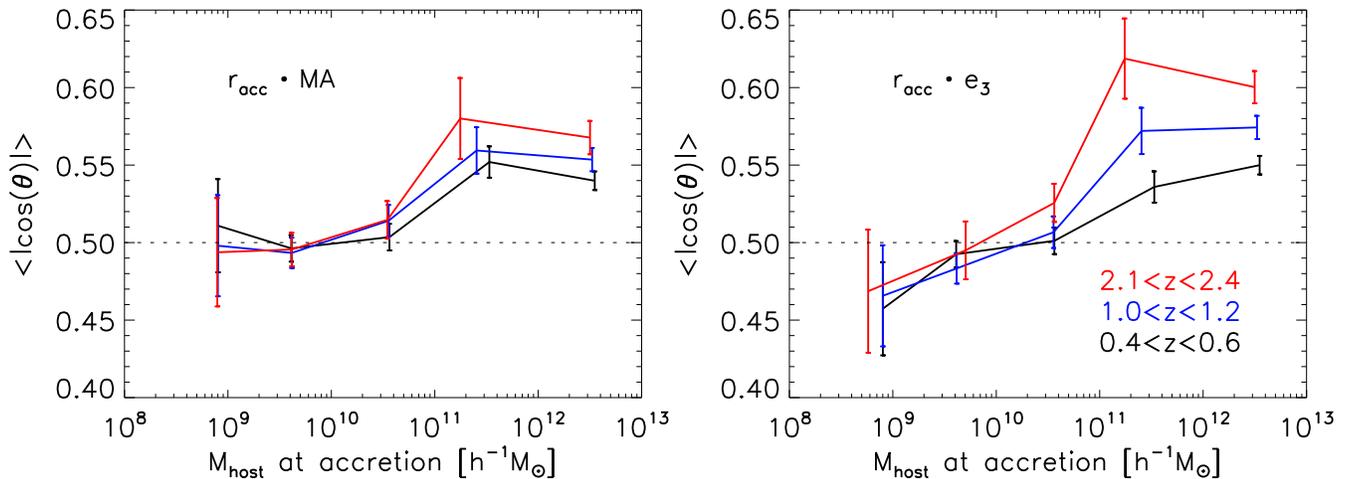,width=0.98\textwidth}}
\caption{The alignment of accreted subhaloes  with the host halo major
  axis (left  panel) and the  $e_{3}$ direction of the  LSS. Different
  from previous  figures, here no  constraints are posed on  the final
  host halo mass, and the accretion angles are binned depending on the
  accretion events with  different host halo mass and  redshift. As we
  can see,  the accretion is  always along  the halo major  axis (left
  panel).  The alignment  with  $e_{3}$  is not  universal  and it  is
  perpendicular to $e_{3}$  for low-mass haloes, and the  flip mass is
  lower at high redshift (right panel).}
\label{fig:infall4}
\end{figure*}

In Fig.~\ref{fig:infall1} the distribution of accretion angle  is  a pure number counting,  and not 
weighted by  the mass  of subhaloes. In Fig.~\ref{fig:infall3} we  plot the  cumulative fraction  of mass accreted along the $e_{3}$ direction (left panel) and along the major
axes of haloes (right panel). Note  that in our analysis  the minimum
halo mass  counts only 20 particles,  so we neglect the  smooth mass
accretion  and  consider only  the   resolved  mass   in  subhaloes.  It is  seen that  in massive
haloes  ($>10^{14}M_{\odot}$), the  net  mass accretion  is along  the
major axis of  the host or along $e_{3}$, and the latter is slightly stronger. 
For  example, more than 50\% of  the  resolved  mass  is accreted  with  $\theta  <30^{\circ}$.  For low-mass host haloes, the mass accretion is close  to be isotropic.

\subsection{Dependence on accretion mass and redshift}
\label{sec:dependence}

Above we have shown the overall  accretion of subhaloes along the halo
major axis and $e_{3}$ of the  LSS.  However, in that analysis we fix
  the  host halo  mass  at  $z=0$ and  study  the  accretion of  their
  progenitors  at high  redshift.   Doing that  is  to understand  the
  primordial alignment of accreted subhaloes and to compare with their
  current  alignment in  the  $z=0$  hosts.  In  this  case, the  main
  progenitors  and the  accreted  subhalo mass  are  all evolved  with
  redshift. In this part, we do not  fix the $z=0$ host, but study the
  alignment for fixed  host halo mass at different  redshift.  In this
  way, we can  quantify the mass dependence and being  able to compare
  with the recent study by Libeskind et al. (2014).

In Fig.~\ref{fig:infall4}  we show the  alignment along the  host halo
major axis (left  panel) and the $e_{3}$ direction  (right panel). The
left panel  shows that  the alignment with  halo major  axis increases
with  host  halo  mass,  and  it   is  stronger  for  haloes  at  high
redshift. Note that here  we do not  scale the  halo mass by  the characteristic
mass    of    halo    collapsing    ($M_{*}$),   as    we    did    in
Fig.~\ref{fig:haloMA-e3}. So for given halo mass, the density contrast
is  higher  at  high  redshift.   Note that  for  massive  haloes  the
alignment  from the  simulation  of $200Mpc/h$  box is  systematically
lower than that from the small box simulation, and this is due to some
low-mass  subhaloes  that have  not  been  resolved in  the  large-box
simulation. Basically, our results  agree with previous studies (e.g.,
Benson 2005; Wang et  al. 2005; Shi et al. 2015)  which state that the
alignment with the halo major axis  is stronger for massive haloes. It
is also seen that the alignment is universal, i.e., that subhaloes are
always  accreted along  the  major axis  although it  is  close to  be
isotropic for low-mass haloes.

The  results in  the  right  panel are  more  interesting. First,  the
accretion with $e_{3}$ of the LSS  is stronger than the alignment with
the halo  major axis shown  in the left  panel, and it  also increases
with host halo  mass and redshift. Moreover, such  dependences on halo
mass and accretion time are much  stronger than the alignment with the
halo  major axis.   Second, it  is seen  that the  alignment with  the
$e_{3}$  is  not  universal.  For low-mass  haloes,  th  accretion  is
actually  perpendicular  to  $e_{3}$,  and the  transit  mass  of  the
alignment is  around $3  \times 10^{10}M_{\odot}/h$  at $z=0$,  and it
decreases to $5\times 10^{9}M_{\odot}/h$ at $z=2$.

 Libeskind et  al. (2014) also studied the  alignment of subhaloes
  with the $e_{3}$ axis of the  LSS around host haloes. However, their
  approaches are different  from ours. Firstly, they  use the velocity
  shear  to  define  LSS,  whereas  we  use  the  density  field  (see
  Forero-Romero et  al. (2014)  for discussion and  comparison between
  the  classification  of  the  LSS  based  on  density  and  velocity
  field). Secondly, the  alignment of subhaloes in  their analysis are
  measured  when subhaloes  cross  a  few virial  radius  ($4 \sim  16
  R_{vir}$)  of  the host  halo.  Although  the approaches  are  quite
  different, we get similar dependences on host halo mass and redshift
  compared with their results. The  main difference is that they found
  that the accretion  is always along $e_{3}$, while we  find that for
  very low mass  haloes, the accretion of  subhaloes are perpendicular
  to $e_{3}$.  Regardness of the  different approached, there  are two
  possible   additional   reasons   for  the   difference   with   our
  results. Firstly, In Libeskind et al. (2014) they combine the signal
  for  host   halo  mass   normalized  by  the   characteristic  mass,
  $M_{\ast}$, of halo collapse.  At $z=0$, $M_{\ast}$ is about $\sim 3
  \times  10^{12}M_{\odot}$.  They  found  that for  all  haloes  with
  $m<0.1M_{\ast}$  the  accretion  is   along  $e_{3}$.  For  a  rough
  comparison, if we also combine  the accretion signal for haloes with
  $M_{host} <0.1M_{\ast}$ at $z=0$, we also find that the alignment is
  along $e_{3}$.  Secondly, the smooth  length in Libeskind et  al. is
  not constant for all haloes, but  it is related to the virial radius
  of each  halo. As we  see in Fig.~\ref{fig:haloMA-e3} using  a small
  smooth length produces  a stronger alignment between  the halo major
  axis  and  the orientation  of  density  field  around, thus  it  is
  expected that the  alignment with the LSS is close  to the alignment
  with the halo major axis, and being more universal.

\section{Conclusion and Discussion}
\label{sec:con}

In this paper, we use N-body simulations to study the anisotropy of subhalo accretion with respect to the halo shape (major axis) and the large scale structure around ($e_{3}$, the least compressed direction). We have obtained these  main results:
\begin{itemize}
\item The major axis of dark matter halo is well aligned with the least compressed direction, $e_{3}$, of the large scale structure around the halo, in agreement with the expectation from the Zel'dovich theory.

\item Subhaloes are accreted along the major axis of the host halo, and the alignment increases with host halo mass and redshift (stronger at high redshift). Most importantly, this alignment is universal (positive) across all halo mass.

\item The accretion of subhalo along $e_{3}$ is not universal. In massive host haloes, the accretion is aligned with $e_{3}$, while in low-mass host haloes, the accretion of subhaloes is perpendicular to $e_{3}$. The transit mass for this flip is lower at high redshift.
\end{itemize}

Our results show that there is a primordial anisotropy of subhaloes at
accretion, and the  degree of alignment is stronger  than the observed
alignment of satellites galaxies at $z=0$. Thus, it is still difficult
to conclude  how much of the  observed signal of galaxies  is from the
primordial alignment or  from the evolution effect  inside dark matter
haloes. In  any case,  the misalignment  between central  galaxies and
host  haloes  is needed  so  as  to  explain  the observed  degree  of
alignment between satellites and the  shape of central galaxies (e.g.,
Dong et al. 2014).

 Our  result about  the accretion with  the $e_{3}$  direction has
  important implications on the correlation  between the halo spin and
  the the LSS. Previous studies  using N-body simulations (e.g., Zhang
  et  al.  2015  and   references  therein)  have  reached  convergent
  conclusion that  the relation between halo  spin and the LSS  is not
  universal. For massive haloes, their  spin is parallel with $e_{3}$,
  while  for   low-mass  haloes,   their  spin  is   perpendicular  to
  $e_{3}$. The  characteristic mass for  the flip of  this correlation
  decreases    with   increasing    redshift   (e.g.,    Trowland   et
  al. 2013).  Some studies  (Codis et  al. 2012;  Welker et  al. 2014;
  Aragon-Calvo et al.  2014; Codis et al. 2015) pointed  out that this
  is  due to  the  disparity  in the  merger  history  of haloes  with
  different  mass.  They speculated  that  for  massive haloes,  their
  subhaloes are accreted along the  filament, and for low-mass haloes,
  their subhaloes  are accreted  perpendicular to the  filament. Under
  the  assumption  that  the  orbital  angular  momentum  of  accreted
  subhaloes is transferred into the host halo spin, it is naturally to
  expected    a   non-universality    of    the    halo   spin    with
  filaments. However, such a spectulation for mass accretion disparity
  is not clearly  shown in their analysis. Our results,  for the first
  time, show that the accretion along $e_{3}$ is really not universal,
  and  the  redshift dependence  also  agrees  with the  results  from
  simulations.

 In deriving the above non-universality of the alignment between subhalo accretion angle with the orientation of the LSS ($e_{3}$), we use a constant smooth length for all haloes (although being evolving with reshift). Most previous results (see Forero-Romero et al. 2014 for references) used a constant smooth length (ranging between $0.5-2$Mpc, well covered by our choice of smooth length), and most of them (if not all) reached agreement on the non-universality of the halo spin-LSS correlation. It implies that by using a constant smooth length for all halo mass, the non-universality on the accretion angle of subhaloes must be in place so as to explain the simulation results. Our results do support this implication. Nevertheless, our choice of smooth length means that we are actually looking at the mass accretion pattern on a fixed scale at given redshift. As shown by some studies (e.g., Aragon-Calvo et al. 2007a) using the multi-scale morphology filter to the density field, and they also found that the halo spin-LSS correlation has a mass dependence. This implies that the anisotropic accretion of subhalo could be both mass and redshift dependent, and it calls for a more complete study on subhalo accretion using different smooth length.

In addition to the subhalo accretion anisotropy, to further understand in detail the correlation between halo spin and LSS, one  needs to study  the formation history of the halo spin in different environment of the LSS, in order to identify which kind of mergers contribute to the final correlation. Such an analysis should make use of the velocity and spin information from N-body simulations. Also, one has to understand why the correlation between subhalo accretion and $e_{3}$ of LSS is different for high and low mass haloes. We will investigate these issues in a future paper.

\section{Acknowledgments}
We thank the anonymous referee for useful comments and suggestions. We also thank Emanuele Contini for careful reading of the manuscript and Noam Libeskind for discussions. This work is supported by the NSF of Jiangsu Province (No. BK20140050), 973 program (No. 2015CB857003, 2013CB834900), the NSFC (No. 11333008) and the Strategic Priority Research Program the emergence of cosmological structures of the Chinese Academy of Science (No. XDB09000000). The  simulations are run  on the Supercomputing  center of CAS.



\begin{thebibliography}{}

 


\bibitem[\protect\citeauthoryear{Arag{\'o}n-Calvo et 
al.}{2007}]{2007ApJ...655L...5A} Arag{\'o}n-Calvo M.~A., van de Weygaert 
R., Jones B.~J.~T., van der Hulst J.~M., 2007a, ApJ, 655, L5 

\bibitem[\protect\citeauthoryear{Arag{\'o}n-Calvo et 
al.}{2007}]{2007AA...474...315} Arag{\'o}n-Calvo M.~A.,Jones B.~J.~T., van de Weygaert R.,  van der Hulst J.~M., 2007b, A\&A, 474, 315 


\bibitem[\protect\citeauthoryear{Arag{\'o}n-Calvo 
\& Yang}{2014}]{2014MNRAS.440L..46A} Arag{\'o}n-Calvo M.~A., Yang L.~F., 2014, MNRAS, 440, L46 


\bibitem[\protect\citeauthoryear{Bailin 
\& Steinmetz}{2005}]{2005ApJ...627..647B} Bailin J., Steinmetz M., 2005, ApJ, 627, 647 


\bibitem[\protect\citeauthoryear{Benson}{2005}]{2005MNRAS.358..551B} Benson 
A.~J., 2005, MNRAS, 358, 551 

\bibitem[\protect\citeauthoryear{Bond}{1996}]{1996Nature.380..603B}Bond J.R., Kofman L., Pogosyan D., 1996, Nature, 380, 603

\bibitem[\protect\citeauthoryear{Bond \& myers}{1996}]{1996ApJS..103....1B}Bond J.R. \& Myers S.T., 1996, ApJS, 103, 1

\bibitem[\protect\citeauthoryear{Brainerd}{2005}]{2005ApJ...628L.101B} 
Brainerd T.~G., 2005, ApJ, 628, L101 


\bibitem[\protect\citeauthoryear{Bryan 
\& Norman}{1998}]{1998ApJ...495...80B} Bryan G.~L., Norman M.~L., 1998, ApJ, 495, 80 


\bibitem[\protect\citeauthoryear{Cautun et al.}{2014}]{2014MNRAS.441.2923C} 
Cautun M., van de Weygaert R., Jones B.~J.~T., Frenk C.~S., 2014, MNRAS, 
441, 2923 


\bibitem[\protect\citeauthoryear{Codis et al.}{2012}]{2012MNRAS.427.3320C} 
Codis S., Pichon C., Devriendt J., Slyz A., Pogosyan D., Dubois Y., Sousbie 
T., 2012, MNRAS, 427, 3320 

\bibitem[\protect\citeauthoryear{Codis et al.}{2015}]{2015MNRAS.452.3369}Codis S., Pichon C., Pogosyan D., 2015, MNRAS, 452, 3369

\bibitem[\protect\citeauthoryear{Debattista et 
al.}{2015}]{2015MNRAS.452.4094D} Debattista V.~P., van den Bosch F.~C., 
Ro{\v s}kar R., Quinn T., Moore B., Cole D.~R., 2015, MNRAS, 452, 4094 


\bibitem[\protect\citeauthoryear{Dong et al.}{2014}]{2014ApJ...791L..33D} 
Dong X.~C., Lin W.~P., Kang X., Ocean Wang Y., Dutton A.~A., Macci{\`o} 
A.~V., 2014, ApJ, 791, L33 


\bibitem[\protect\citeauthoryear{Dubois et al.}{2014}]{2014MNRAS.444.1453D} 
Dubois Y., et al., 2014, MNRAS, 444, 1453 

\bibitem[\protect\citeauthoryear{Eisenstein \& Loeb}{1996}]{1996ApJ...459..432}Eisenstein D.J. \& Loeb A., 1996, ApJ, 459, 432

\bibitem[\protect\citeauthoryear{Faltenbacher et 
al.}{2007}]{2007ApJ...662L..71F} Faltenbacher A., Li C., Mao S., van den 
Bosch F.~C., Yang X., Jing Y.~P., Pasquali A., Mo H.~J., 2007, ApJ, 662, 
L71 


\bibitem[\protect\citeauthoryear{Forero-Romero, Contreras, 
\& Padilla}{2014}]{2014MNRAS.443.1090F} Forero-Romero J.~E., Contreras S., Padilla N., 2014, MNRAS, 443, 1090 


\bibitem[\protect\citeauthoryear{Hahn et al.}{2007a}]{2007MNRAS.375..489H} 
Hahn O., Porciani C., Carollo C.~M., Dekel A., 2007, MNRAS, 375, 489 

\bibitem[\protect\citeauthoryear{Hahn et al.}{2007b}]{2007MNRAS.381..41H} 
Hahn O., Carollo C.~M., Porciani C., Dekel A., 2007, MNRAS, 381, 41

\bibitem[\protect\citeauthoryear{Ibata et al.}{2013}]{2013Natur.493...62I} 
Ibata R.~A., et al., 2013, Nature, 493, 62 

\bibitem[\protect\citeauthoryear{Icke}{1973}]{1973AA...27....1}Icke V., 1973, A\&A, 27, 1

\bibitem[\protect\citeauthoryear{Jing 
\& Suto}{2002}]{2002ApJ...574..538J} Jing Y.~P., Suto Y., 2002, ApJ, 574, 538 


\bibitem[\protect\citeauthoryear{Joachimi et 
al.}{2015}]{2015SSRv..tmp...65J} Joachimi B., et al., 2015, SSRv, 65 


\bibitem[\protect\citeauthoryear{Jones, van de Weygaert, 
\& Arag{\'o}n-Calvo}{2010}]{2010MNRAS.408..897J} Jones B.~J.~T., van de Weygaert R., Arag{\'o}n-Calvo M.~A., 2010, MNRAS, 408, 897 


\bibitem[\protect\citeauthoryear{Kang et al.}{2005}]{2005ApJ...631...21K} 
Kang X., Jing Y.~P., Mo H.~J., B{\"o}rner G., 2005, ApJ, 631, 21 


\bibitem[\protect\citeauthoryear{Kang et al.}{2007}]{2007MNRAS.378.1531K} 
Kang X., van den Bosch F.~C., Yang X., Mao S., Mo H.~J., Li C., Jing Y.~P., 
2007, MNRAS, 378, 1531 


\bibitem[\protect\citeauthoryear{Komatsu et 
al.}{2011}]{2011ApJS..192...18K} Komatsu E., et al., 2011, ApJS, 192, 18 


\bibitem[\protect\citeauthoryear{Kroupa, Theis, 
\& Boily}{2005}]{2005A&A...431..517K} Kroupa P., Theis C., Boily C.~M., 2005, A\&A, 431, 517 


\bibitem[\protect\citeauthoryear{Lee}{2004}]{2004ApJ...614L...1L} Lee J., 
2004, ApJ, 614, L1 


\bibitem[\protect\citeauthoryear{Lee 
\& Pen}{2000}]{2000ApJ...532L...5L} Lee J., Pen U.-L., 2000, ApJ, 532, L5 


\bibitem[\protect\citeauthoryear{Libeskind et 
al.}{2005}]{2005MNRAS.363..146L} Libeskind N.~I., Frenk C.~S., Cole S., 
Helly J.~C., Jenkins A., Navarro J.~F., Power C., 2005, MNRAS, 363, 146 


\bibitem[\protect\citeauthoryear{Libeskind et 
al.}{2012}]{2012MNRAS.421L.137L} Libeskind N.~I., Hoffman Y., Knebe A., 
Steinmetz M., Gottl{\"o}ber S., Metuki O., Yepes G., 2012, MNRAS, 421, L137 


\bibitem[\protect\citeauthoryear{Libeskind et 
al.}{2014}]{2014MNRAS.443.1274L} Libeskind N.~I., Knebe A., Hoffman Y., 
Gottl{\"o}ber S., 2014, MNRAS, 443, 1274 


\bibitem[\protect\citeauthoryear{Metz, Kroupa, 
\& Jerjen}{2007}]{2007MNRAS.374.1125M} Metz M., Kroupa P., Jerjen H., 2007, MNRAS, 374, 1125 


\bibitem[\protect\citeauthoryear{Okumura, Jing, 
\& Li}{2009}]{2009ApJ...694..214O} Okumura T., Jing Y.~P., Li C., 2009, ApJ, 694, 214 


\bibitem[\protect\citeauthoryear{Porciani, Dekel, 
\& Hoffman}{2002}]{2002MNRAS.332..325P} Porciani C., Dekel A., Hoffman Y., 2002, MNRAS, 332, 325 

\bibitem[\protect\citeauthoryear{Rossi}{2013}]{2013MNRAS.430.1486}Rossi G., 2013, MNRAS, 430, 1486

\bibitem[\protect\citeauthoryear{Shi, Wang, 
\& Mo}{2015}]{2015ApJ...807...37S} Shi J., Wang H., Mo H.~J., 2015, ApJ, 807, 37 
\bibitem[\protect\citeauthoryear{Silk White}{1978}]{1978ApJ...223L..59S}Silk J. \& White S.D.M., 1978, ApJ, 223, L59

\bibitem[\protect\citeauthoryear{Springel}{2005}]{2005MNRAS.364.1105S} 
Springel V., 2005, MNRAS, 364, 1105 


\bibitem[\protect\citeauthoryear{Springel, Frenk, 
\& White}{2006}]{2006Natur.440.1137S} Springel V., Frenk C.~S., White S.~D.~M., 2006, Natur, 440, 1137 


\bibitem[\protect\citeauthoryear{Springel et 
al.}{2001}]{2001MNRAS.328..726S} Springel V., White S.~D.~M., Tormen G., 
Kauffmann G., 2001, MNRAS, 328, 726 


\bibitem[\protect\citeauthoryear{Tempel 
\& Libeskind}{2013}]{2013ApJ...775L..42T} Tempel E., Libeskind N.~I., 2013, ApJ, 775, L42 


\bibitem[\protect\citeauthoryear{Trowland, Lewis, 
\& Bland-Hawthorn}{2013}]{2013ApJ...762...72T} Trowland H.~E., Lewis G.~F., Bland-Hawthorn J., 2013, ApJ, 762, 72 


\bibitem[\protect\citeauthoryear{Trujillo, Carretero, 
\& Patiri}{2006}]{2006ApJ...640L.111T} Trujillo I., Carretero C., Patiri S.~G., 2006, ApJ, 640, L111 



\bibitem[\protect\citeauthoryear{van de Weygaert \& }{1996}]{1996MNRAS...281..84V} van de Weygaert R. \& Bertschinger E., 1996, MNRAS, 281, 84 


\bibitem[\protect\citeauthoryear{van Haarlem 
\& van de Weygaert}{1993}]{1993ApJ...418..544V} van Haarlem M., van de Weygaert R., 1993, ApJ, 418, 544 


\bibitem[\protect\citeauthoryear{Vogelsberger et 
al.}{2014}]{2014MNRAS.444.1518V} Vogelsberger M., et al., 2014, MNRAS, 444, 
1518 


\bibitem[\protect\citeauthoryear{Wang et al.}{2005}]{2005MNRAS.364..424W} 
Wang H.~Y., Jing Y.~P., Mao S., Kang X., 2005, MNRAS, 364, 424 


\bibitem[\protect\citeauthoryear{Wang et al.}{2014}]{2014ApJ...786....8W} 
Wang Y.~O., Lin W.~P., Kang X., Dutton A., Yu Y., Macci{\`o} A.~V., 2014, 
ApJ, 786, 8 

\bibitem[\protect\citeauthoryear{Welker et al.}{2014}]{2014MNRAS.445..46}
Welker C., Devriendt J., Dubois Y., Pichon C., Peirani S., 2014, MNRAS, 445, 46

\bibitem[\protect\citeauthoryear{White}{1984}]{1984ApJ...286...38W} White 
S.~D.~M., 1984, ApJ, 286, 38 


\bibitem[\protect\citeauthoryear{Yang et al.}{2006}]{2006MNRAS.369.1293Y} 
Yang X., van den Bosch F.~C., Mo H.~J., Mao S., Kang X., Weinmann S.~M., 
Guo Y., Jing Y.~P., 2006, MNRAS, 369, 1293 


\bibitem[\protect\citeauthoryear{Zel'dovich}{1970}]{1970A&A.....5...84Z} Zel'dovich Y.~B., 1970, A\&A, 5, 84 


\bibitem[\protect\citeauthoryear{Zhang et al.}{2009}]{2009ApJ...706..747Z} 
Zhang Y., Yang X., Faltenbacher A., Springel V., Lin W., Wang H., 2009, 
ApJ, 706, 747 


\bibitem[\protect\citeauthoryear{Zhang et al.}{2015}]{2015ApJ...798...17Z} 
Zhang Y., Yang X., Wang H., Wang L., Luo W., Mo H.~J., van den Bosch F.~C., 
2015, ApJ, 798, 17 


\end{thebibliography}
\end{document}